\documentclass{aastex63}
\usepackage{amsmath}
\usepackage{subfigure}
\usepackage{threeparttable}
\graphicspath{{./}{figures/}}

\begin{document}

\title{Revised the $\gamma$-ray emission from SNR CTB 109 with {\em Fermi}-LAT}

\author[0000-0001-5135-5942]{Yuliang Xin}
\author{Qizhen Zhao}
\author{Xiaolei Guo}
\affiliation{School of Physical Science and Technology, Southwest Jiaotong University, Chengdu 610031, China; \\
\href{mailto:ylxin@swjtu.edu.cn}{ylxin@swjtu.edu.cn}; \href{mailto:xlguo@swjtu.edu.cn}{xlguo@swjtu.edu.cn}}




\begin{abstract}

CTB 109 is a middle-aged shell-type SNR with bright thermal X-ray emission.
We reanalyze the GeV $\gamma$-ray emission from CTB 109 using thirteen years of Pass 8 data recorded by the {\em Fermi} Large Area Telescope ({\em Fermi}-LAT).
The $\gamma$-ray emission of CTB 109 shows a center bright morphology, which is well consistent with its thermal X-ray emission rather than the shell-type structure in the radio band.
The spectral analysis shows an evident spectral curvature at $\sim$ several GeV for the GeV $\gamma$-ray spectrum, which can naturally explain the lack of TeV $\gamma$-ray emission from CTB 109.
Although either a leptonic or a hadronic model could fit the multi-wavelength observations of CTB 109,
the hadronic model is favored considering its $\gamma$-ray morphology and the spectral curvature of GeV spectrum.
The unusual $\gamma$-ray spectrum of CTB 109 with other SNRs and the luminosity-diameter squared relation make CTB 109 to be distinguished both from the young-aged SNRs with hard GeV $\gamma$-ray spectra and several old-aged SNRs interacting with molecular clouds.

\end{abstract}


\section{Introduction}
\label{intro}

Supernova remnants (SNRs) are widely believed to be the dominated accelerators of Galactic cosmic rays (CRs) with energies up to the knee.
Radio and non-thermal X-ray emission detected provide the clear evidence of electrons acceleration in SNRs \citep{1995Natur.378..255K, 2007Natur.449..576U}.
{\em Fermi} Large Area Telescope ({\em Fermi}-LAT) detected the characteristic ``$\pi^0$ bump'' from several old-aged $\gamma$-ray SNRs interacting with ambient molecular clouds \citep[MCs;][]{2013Sci...339..807A, 2016ApJ...816..100J}, which has been considered as the most direct evidence of nuclei acceleration in SNRs.
The $\gamma$-ray emission from such SNRs are from the decay of neutral pions due to the inelastic nuclei-nuclei collisions, which is the hadronic process.
In addition, the $\gamma$-ray emission can also be produced by leptonic process, that is the inverse Compton scattering or the bremsstrahlung process of relativistic electrons.
And there has been an on-going debate for the origin of the $\gamma$-ray emission from most SNRs \citep{2018A&A...612A...6H, 2019ApJ...874...50Z, 2021ApJ...910...78Z}.

CTB 109 (G109.1-1.0) is a middle-aged shell-type SNR \citep[$\sim$ 9 kyr;][]{2018MNRAS.473.1705S}, which was first discovered in X-ray with the observations of Einstein satellite \citep{1980Natur.287..805G}, and later in radio with the Westerbork Synthesis Radio Telescope \citep{1981ApJ...246L.127H}.
The morphology of CTB 109 shows a semicircular shell in both radio and X-ray observations, which is widely explained by the interaction between the shock of SNR and a neighboring giant molecular cloud \citep{1987PASJ...39..755T,2002ApJ...576..169K}.
The X-ray emission from the remnant is total thermal, without evidence for the synchrotron non-thermal component \citep{1997ApJ...484..828R, 1998A&A...330..175P, 1998AdSpR..22.1039R, 1999NuPhS..69...96P}.
Based on the analysis of HI absorption and CO emission spectra of CTB 109, the distance of it is estimated to be 3 - 4 kpc \citep{2002ApJ...576..169K,2010MNRAS.404L...1T,2012ApJ...746L...4K}.
And the latest measurement by \citet{2018MNRAS.473.1705S} determined the value of 3.1 $\pm$ 0.2 kpc. Hereafter, we adopt this value as the distance of CTB 109.

In the center of CTB 109, an X-ray point source was first detected by Einstein satellite \citep{1980Natur.287..805G}, and then identified to be an anomalous X-ray pulsar \citep[AXP J2301+5852; also 1E 2259+589;][] {1981Natur.293..202F,2004ApJ...605..378W,2004ApJ...607..959G}, which is one type of magnetars.
The measured period of 1E 2259+589 is 6.97 s \citep{1988ApJ...333..777M}, with a magnetic field strength of 5.9 $\times$ 10$^{13}$ G \citep{2013ApJ...772...31T}.
The detection of 1E 2259+589 makes CTB 109 to be one of few SNRs that hosts a magnetar within its boundary.
However, there is no evidence for the component of pulsar wind nebula (PWN) been detected around 1E 2259+589. 
Moreover, the X-ray observations revealed an X-ray bright interior region in CTB 109, called the X-ray Lobe, which was initially suggested to be a jet associated with AXP 1E 2259+586 \citep{1983IAUS..101..429G}. 
More detailed analysis with {\em XMM}-Newton \citep{2004ApJ...617..322S}, {\em Chandra} \citep{2006ApJ...642L.149S,2013A&A...552A..45S}, and {\em Suzaku} \citep{2015PASJ...67....9N,2017PASJ...69...40N} show the thermal X-ray emission from the Lobe and suggest that there is no morphological connection with the magnetar.

The $\gamma$-ray emission from CTB 109 was first reported by \citet{2012ApJ...756...88C} with 37 months of data from {\em Fermi}-LAT. 
The data analysis shows no significant extension for the $\gamma$-ray morphology of CTB 109.
And its $\gamma$-ray spectrum is hard with a power-law index of $\Gamma$ = 2.07 $\pm$ 0.12, which shows no indications for spectral curvature in the GeV range.
Both the leptonic and hadronic models are discussed to explain the $\gamma$-ray emission from CTB 109, and neither of them can be ruled out \citep{2012ApJ...756...88C}.
In \citet{2019ICRC...36..674F}, HAWC tried to search for the TeV $\gamma$-ray emission from CTB 109. However, there is no significant TeV emission been detected and only the upper limits are given. 

In the present work, we carry out a detailed analysis of the GeV $\gamma$-ray emission from SNR CTB 109 with the latest Pass 8 data recorded by {\em Fermi}-LAT. And this paper is organized as follows. In Section 2, we describe the data analysis routines and results. A discussion of the origin of the $\gamma$-ray emission based on the multi-wavelength observations is given in Section 3, together with the comparison with other SNRs. And the conclusion of this work is presented in Section 4.

\section{{\em Fermi}-LAT Data Analysis}
\label{fermi}

\subsection{Data Reduction}
\label{data reduction}

In following analysis, the latest {\em Fermi}-LAT Pass 8 data are collected from August 4, 2008 (Mission Elapsed Time 239557418) 
to August 4, 2021 (Mission Elapsed Time 649728005). We select the data with ``Source'' event class (evclass = 128 \& evtype=3)
and exclude the events with zenith angles greater than $90^\circ$ to reduce the contamination from Earth Limb.
The energies of photons are cut between 10 GeV and 1 TeV for the spatial analysis considering the impact of large point spread function 
in the lower energy band.
And for the spectral analysis, the events with energy from 100 MeV to 1 TeV are used.
The region of interest (ROI) we analysed is a square region of $14^\circ \times 14^\circ$ 
centered at the position of CTB 109 in the incremental version of the fourth {\em Fermi}-LAT source catalog \citep[4FGL-DR2;][]{2020ApJS..247...33A, 2020arXiv200511208B}. 
The binned analysis method is used and the standard {\it ScienceTools} software package \footnote {http://fermi.gsfc.nasa.gov/ssc/data/analysis/software/}, together with the instrumental response function (IRF) 
of ``P8R3{\_}SOURCE{\_}V3'', are adopted. The Galactic and isotropic diffuse background models used here are
{\tt gll\_iem\_v07.fits} and {\tt iso\_P8R3\_SOURCE\_V3\_v1.txt}
\footnote {http://fermi.gsfc.nasa.gov/ssc/data/access/lat/BackgroundModels.html}, respectively.
All sources listed in the 4FGL-DR2 catalog within a radius of $20^\circ$ from the ROI center together with the two diffuse backgrounds
are included in the source model, generated by the user-contributed software 
{\tt make4FGLxml.py}\footnote{http://fermi.gsfc.nasa.gov/ssc/data/analysis/user/}.
During the fitting procedure, the normalizations and spectral parameters of sources within $7^\circ$ around CTB 109, 
together with the normalizations of the two diffuse backgrounds, are left free.

\subsection{Results}
\label{results}
\begin{figure*}[!htb]
	\centering
    \includegraphics[width=0.32\textwidth]{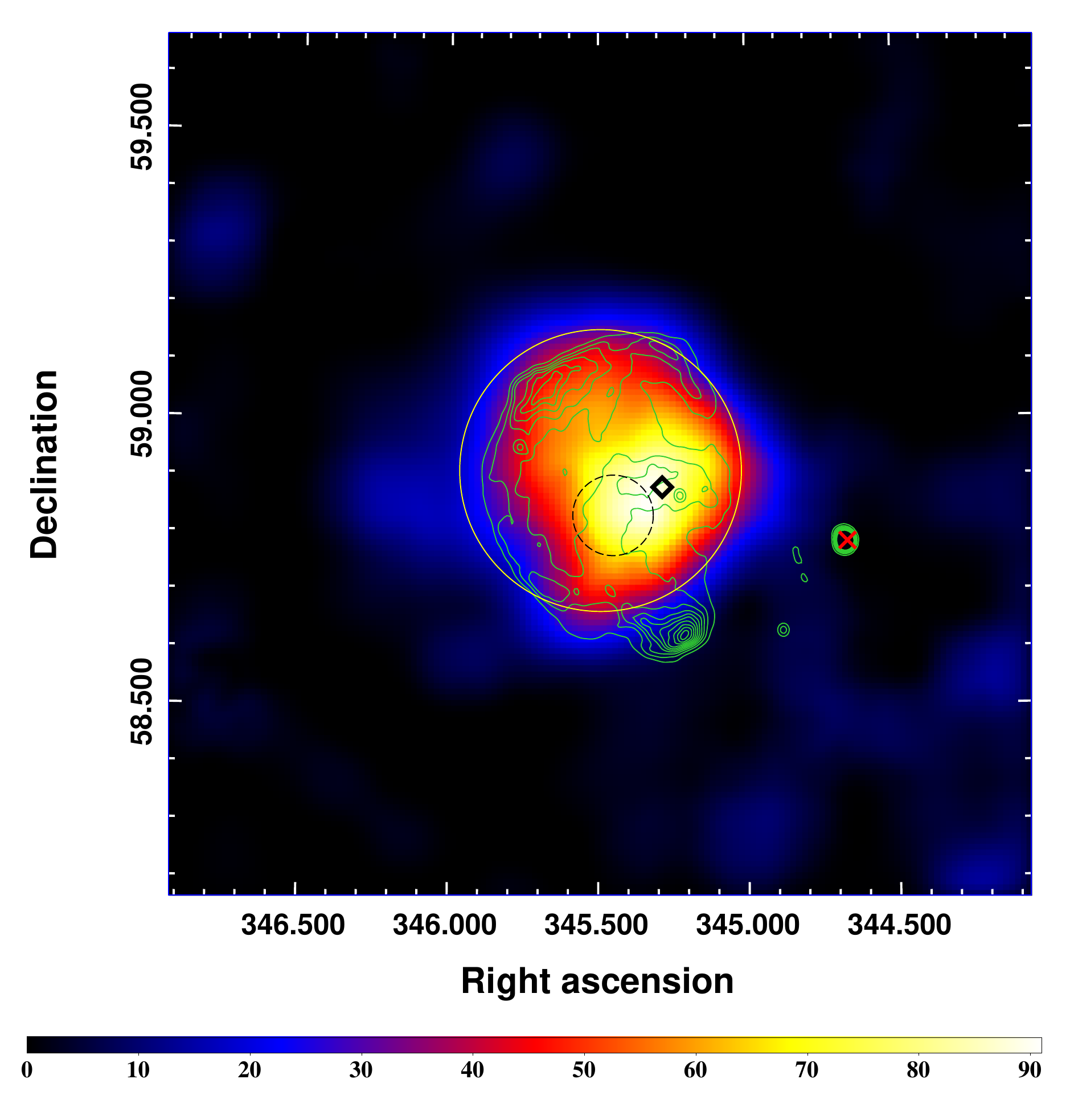}
    \includegraphics[width=0.32\textwidth]{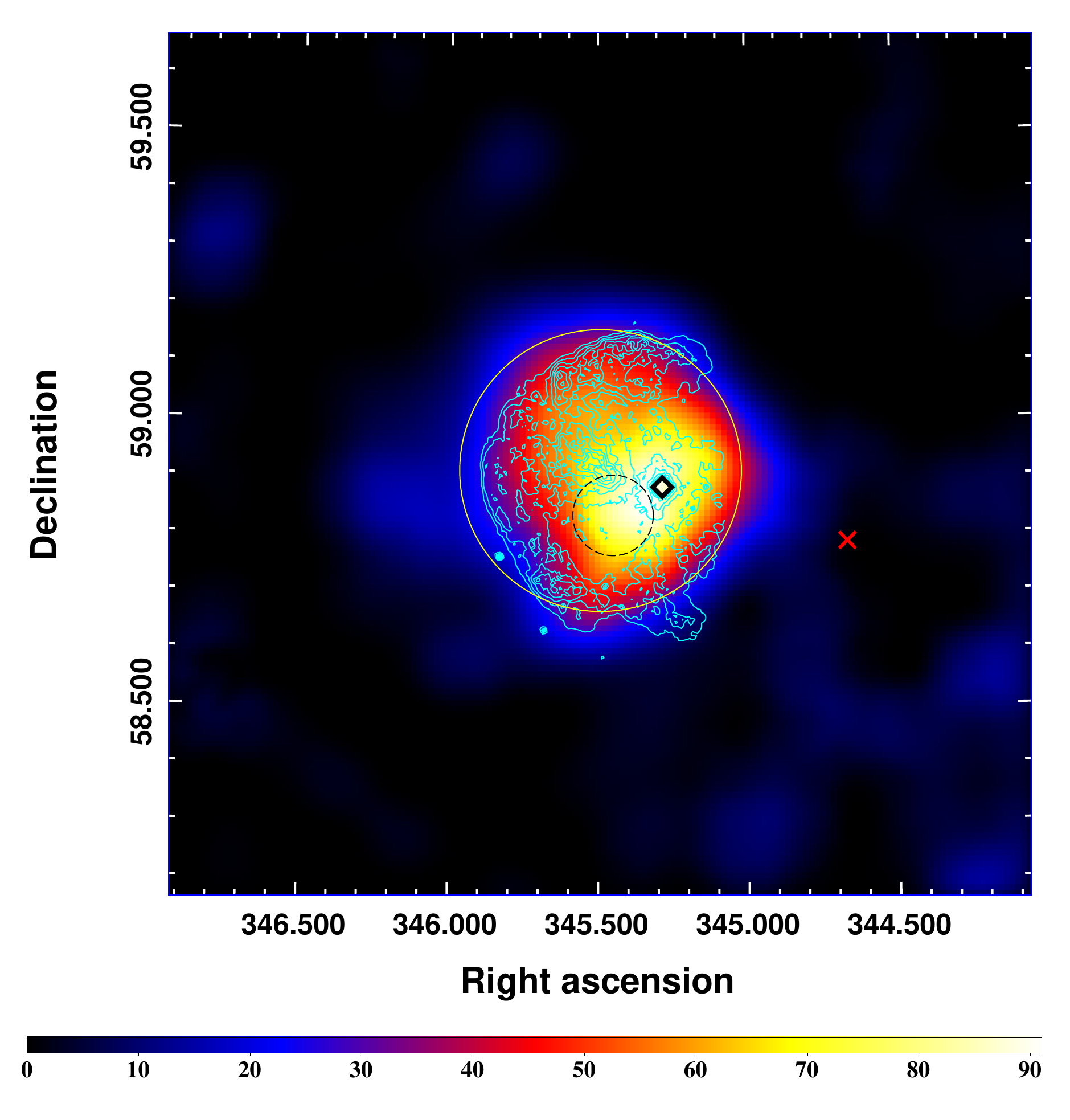}
    \includegraphics[width=0.32\textwidth]{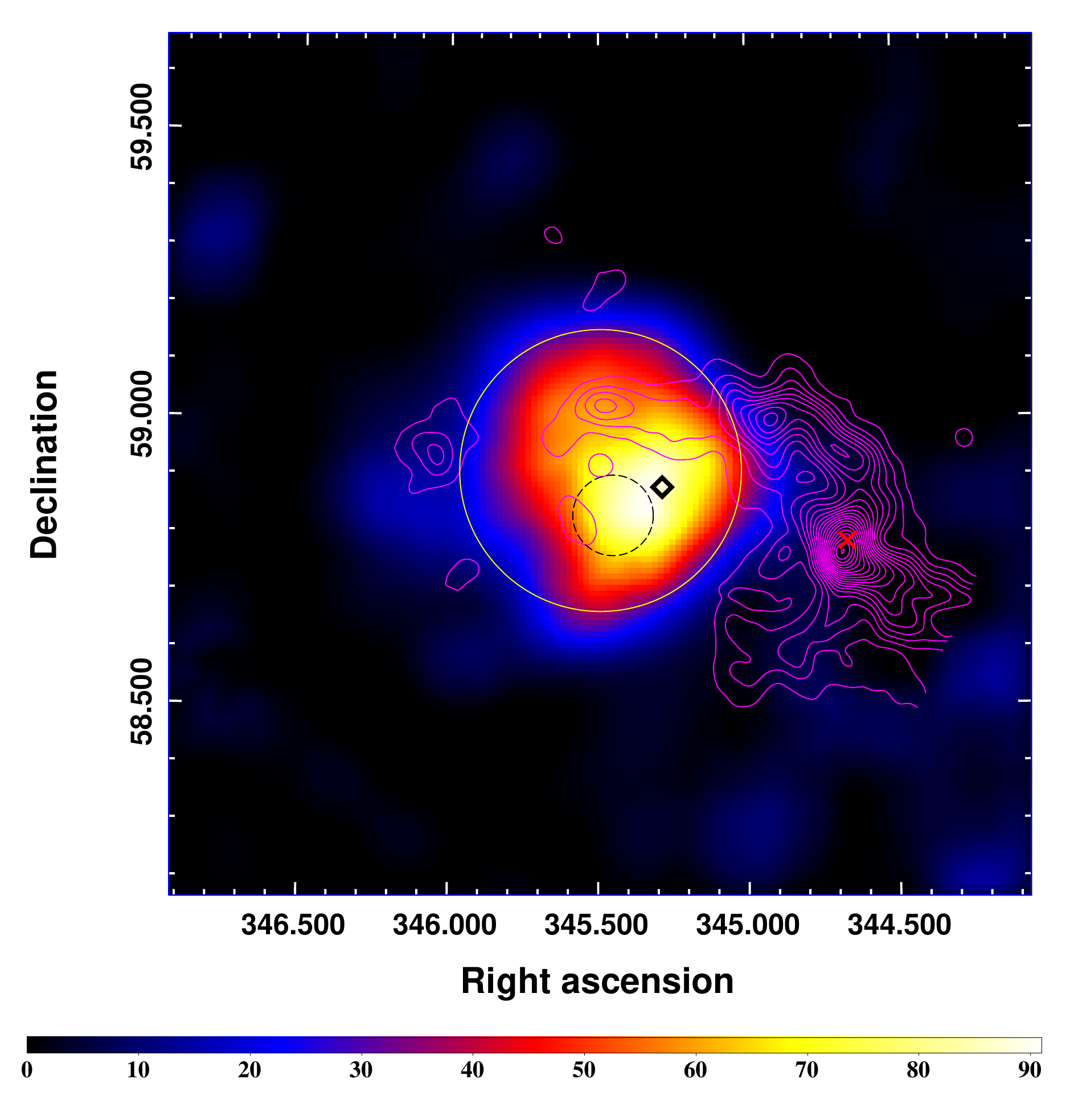}
	\caption{TSmaps of $1^{\circ}\!.5$ $\times$ $1^{\circ}\!.5$ centered at CTB 109 for photons above 10 GeV. These maps are smoothed with a Gaussian kernal of $\sigma$ = $0^{\circ}\!.02$. The yellow sold circle shows the best-fitting radius of the uniform disk for the spatial template of CTB 109, and the black dashed circle marks the centroid position fitted by \cite{2012ApJ...756...88C} as a point source with its 95\% confidence radius. The black diamond reveals the magnetar 1E 2259+589 and the red cross presents the position of 4FGL J2258.6+5847c (Sh 2-152). The green contours(left) represent the radio image of CTB 109 from DRAO at 1.4 GHz \citep{1984ApJ...283..147H}; the cyan contours(middle) show the thermal X-ray emission from {\em ROSAT} HRI \citep{1997ApJ...484..828R} and the magenta contours display the the $^{12}$CO ($J$=1$-$0) emission from FCRAO integrated the velocity between -48 km s$^{-1}$ and -56 km s$^{-1}$ \citep{1998ApJS..115..241H, 2012ApJ...746L...4K}}.
	\label{fig1:tsmap}
\end{figure*}

In the 4FGL-DR2 catalog, SNR CTB 109 corresponds to the extended $\gamma$-ray source 4FGL J2301.9+5855e and the $\gamma$-ray emission is described by an uniform disk with a radius of $0^{\circ}\!.249$ \citep{2020arXiv200511208B}.
In order to show the spatial correlation between the $\gamma$-ray and other energy bands, we created a test statistic (TS) map with photons above 10 GeV shown in Figure \ref{fig1:tsmap}.
And the centroid of the $\gamma$-ray emission is closer to the northwest of SNR than that detected by \cite{2012ApJ...756...88C}.
The radio contours at 1420 MHz from Synthesis Telescope of the Dominion Radio Astrophysical Observatory \citep[DRAO;][]{1984ApJ...283..147H} survey, the X-ray contours from {\em ROSAT} High Resolution Image \citep[HRI;][]{1997ApJ...484..828R} and the $^{12}$CO ($J$=1$-$0) emission from Five College Radio Astronomy Observatory \citep[FCRAO;][]{1998ApJS..115..241H} survey Here the CO intensity is calculated by integrating the velocity between -48 km s$^{-1}$ and -56 km s$^{-1}$, which corresponds to the distance estimation of 3.2$\pm$0.2 kpc for CTB 109 by \citet{2012ApJ...746L...4K}.
The TSmap reveals that the GeV $\gamma$-ray emission from CTB 109 mainly concentrates in the central region of the remnant, which is not very consistent with the morphology of radio shell.

To determine the morphology of the $\gamma$-ray emission from CTB 109, we also used {\tt Fermipy}, 
a {\tt PYTHON} package that automates analyses with the Fermi Science Tools \citep{2017ICRC...35..824W}, to test the size of the uniform disk with the events above 10 GeV.
The best-fitting position and radius of the uniform disk are R.A.$=345^{\circ}\!.492 \pm 0^{\circ}\!.022$, Dec.$=58^{\circ}\!.908 \pm 0^{\circ}\!.018$, and $0^{\circ}\!.245 \pm 0^{\circ}\!.013$, 
respectively, which are consistent with the results in the 4FGL-DR2 catalog \citep{2020arXiv200511208B}.
The TS value of the uniform disk is estimated to be 155.2.
The best-fitting two-dimensional Gaussian template given by {\tt Fermipy} (R.A.$=345^{\circ}\!.459 \pm 0^{\circ}\!.028$, Dec.$=58^{\circ}\!.899 \pm 0^{\circ}\!.024$, $\sigma$ = $0^{\circ}\!.148 \pm 0^{\circ}\!.017$) is also tested, which gives the similar TS value ($\sim$ 154.0) for CTB 109.

With the spatial template of an uniform disk, we perform a spectral analysis of CTB 109 in the energy range from 100 MeV to 1 TeV.
Similar to the analysis procedure of Puppis A in \citet{2017ApJ...843...90X}, we first fit the data using a simple power-law (PL) spectrum.
And the spectral index is fitted to be 1.90 $\pm$ 0.04, which is slightly harder than the result of 2.07 $\pm$ 0.12 in \citet{2012ApJ...756...88C}. And the integral photon flux is calculated to be $(1.15 \pm 0.13)\times 10^{-8}$ photon cm$^{-2}$ s$^{-1}$.
To test the spectral curvature of the $\gamma$-ray emission from CTB 109, a log-parabola model ($dN/dE \propto E^{-(\alpha+\beta {\rm log}(E/E_b))}$) is adopted.
And the best-fitting parameters are shown in Table \ref{table:spectra}.
The higher TS value with a log-parabola model shows an evidence of spectral curvature of CTB 109 ($\sim$3.4$\sigma$ improvement with respect to a power-law model with one additional free model parameter.

\begin{figure*}[!htb]
	\centering
    \includegraphics[width=0.49\textwidth]{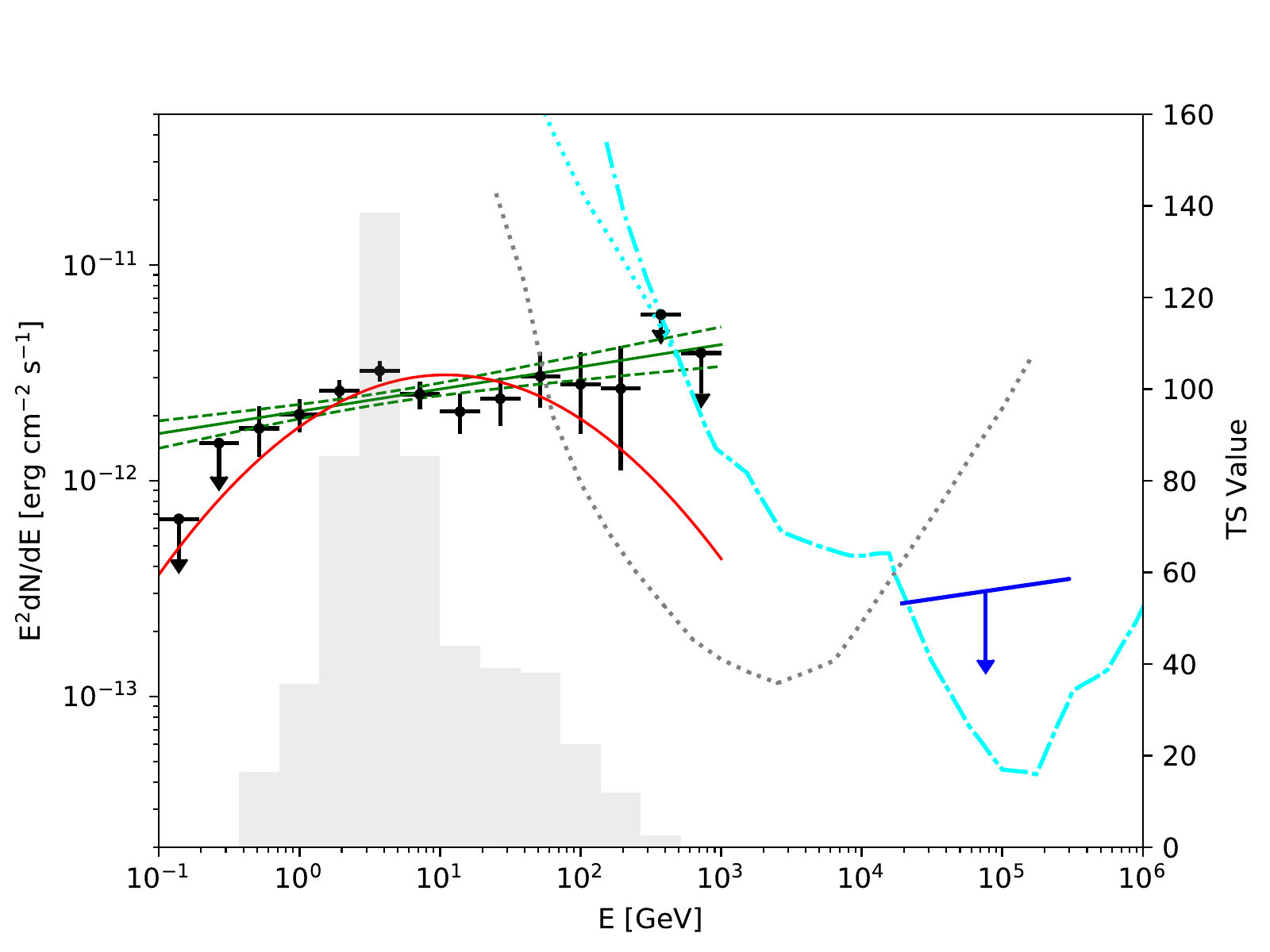}
    \includegraphics[width=0.49\textwidth]{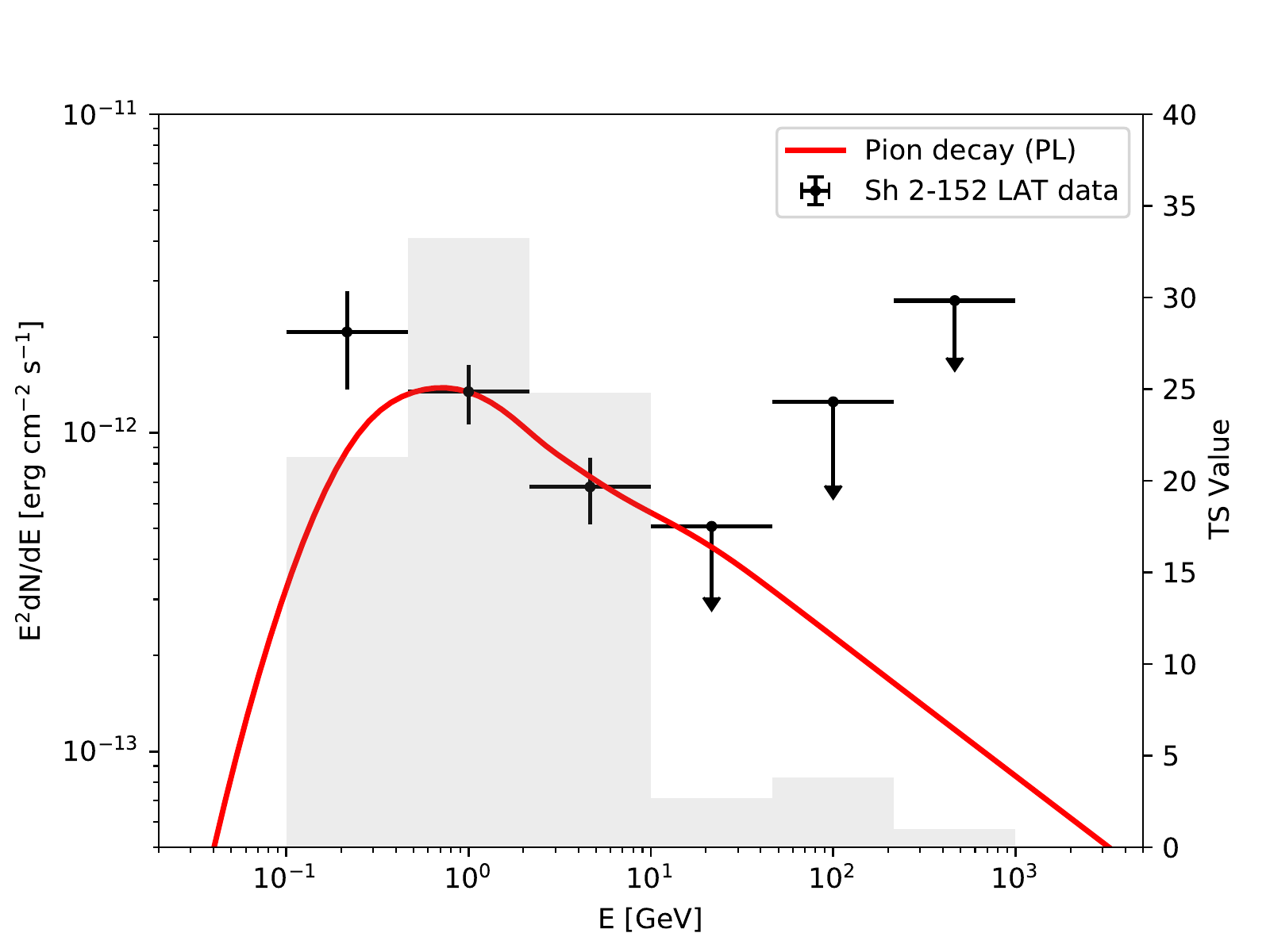}
	\caption{Left: SED of SNR CTB 109. The results of {\em Fermi}-LAT data are shown as the black dots, with arrows indicating the 95\% upper limits. The gray histogram shows the TS value for each energy bin. The TeV upper limit (blue arrow) are from HAWC observation \citep{2019ICRC...36..674F}. The green solid and dashed lines are the best-fitting power-law spectrum and its 1$\sigma$ statistic error  in the energy range of 100 MeV - 1 TeV. The red sold line is the best-fitting log-parabola spectrum. The cyan dotted and dotted–dashed lines represent the differential sensitivities of LHAASO (1 yr) with different sizes of photomultiplier tube \citep[PMT;][]{2019arXiv190502773C}. And the black dotted line shows the differential sensitivity of CTA-North \citep[50 hr;][]{2019scta.book.....C}. Right: SED of Sh 2-152 shown as the black dots. The red line shows the hadronic emission by adopting a power-law spectrum for protons.}
	\label{fig2:sed}
\end{figure*}

Furthermore, to obtain the spectral energy distribution (SED) of CTB 109, we divide the data into 14 logarithmic equal energy bins from 100 MeV to 1 TeV, and perform the same likelihood fitting analysis for each energy bin.
For the bin with TS value of CTB 109 smaller than 5.0, an upper limit with 95\% confidence level is calculated.
The results are presented as the black dots in Figure \ref{fig2:sed}.
The global fitting with the power-law and log-parabola models are also overplotted, together with the upper limit of the TeV $\gamma$-ray flux derived by HAWC \citep{2019ICRC...36..674F}.
And the existence of the spectral curvature of CTB 109 in the GeV band can naturally explain the non-detection in the TeV $\gamma$-ray band.

\begin{table}[!htb]
\caption {Comparison of Spectral Models (100 MeV - 1 TeV)}
\begin{tabular}{ccccccc}
\hline \hline
Model   &  Index$^{\,\,\text{a}}$  &   $\beta$   &  Photon Flux           & Energy Flux        & Number of  & TS Value   \\
        &                          &             &  $\rm 10^{-9}$ $\rm ph$ $\rm cm^{-2}$ $\rm s^{-1}$  &  $\rm 10^{-5}$ $\rm MeV$ $\rm cm^{-2}$ $\rm s^{-1}$ &     free model parameters  &            \\
\hline
Power-Law    & 1.90 $\pm$ 0.04   & --                &    11.52 $\pm$ 1.28     &1.59 $\pm$ 0.13   & 2   & 523.3   \\
Log-parabola & 1.60 $\pm$ 0.10   & 0.086 $\pm$ 0.023 &    6.46 $\pm$ 1.16      &1.10 $\pm$ 0.11   & 3   & 534.6   \\
\hline
\end{tabular}
\begin{tablenotes}
\footnotesize
\item[a] The Index column is for both index (power-law), and for $\alpha$ (log-parabola).
\end{tablenotes}
\label{table:spectra}
\end{table}

In the west of CTB 109, there is a $\gamma$-ray point source, 4FGL J2258.6+5847c, which is associated with the star-forming compact HII region Sh 2-152 \citep{2000A&A...364..723D,2011A&A...535A...8R}.
In the energy band of 100 MeV - 1 TeV, the $\gamma$-ray emission from Sh 2-152 can be described by a power-law model with an index of 2.25 $\pm$ 0.12, and the TS value is fitted to be 67.1.
And the energy flux of the $\gamma$-ray emission is calculated to be $\rm (3.76 \pm 0.62) \times 10^{-6}$ $\rm MeV$ $\rm cm^{-2}$ $\rm s^{-1}$.
Based on the observations of the inside young stellar objects (YSOs), the distance of Sh 2-152 is estimated to be 3.21 $\pm$ 0.21 kpc \citep{2011A&A...535A...8R}.
With this value, the $\gamma$-ray luminosity of Sh 2-152 is calculated to be (7.43 $\pm$ 1.22) $\rm \times 10^{33}$ $\rm erg$ $\rm s^{-1}$.
The resulting SED of Sh 2-152 is shown in the right panel of Figure \ref{fig2:sed}.

\section{Discussion}
\label{discussion}

From the above analysis, the extended GeV $\gamma$-ray emission from CTB 109 is spatially well-correlated with the morphology in the thermal X-ray rather than the radio band.
And the $\gamma$-ray emission shows the center bright morphology, which is not consistent with the shell-type structure in the radio band.
All evidences above support the hadronic scenario for the $\gamma$-ray emission from CTB 109, 
which also means that there should be some dense clouds in the $\gamma$-ray emitting region like SNR Puppis A \citep{2017ApJ...843...90X}.
However, the right panel of Figure \ref{fig1:tsmap} shows no bright CO emission detected in the centroid position of the $\gamma$-ray emission region, except some small surrounding molecular clouds \citep{2006ApJ...642L.149S,2012ApJ...756...88C}.
This could be attributed to dissociation of molecules by the radiative precursor of the SNR due to the photoionization and photodissociation effects \citep{2008A&A...480..439P}.
The absence of CO emission may also stems from the CO-dark gas in the $\gamma$-ray emitting region that cannot be traced by CO observations \citep{2005Sci...307.1292G}.

\begin{figure*}[!htb]
	\centering
    \includegraphics[width=0.49\textwidth]{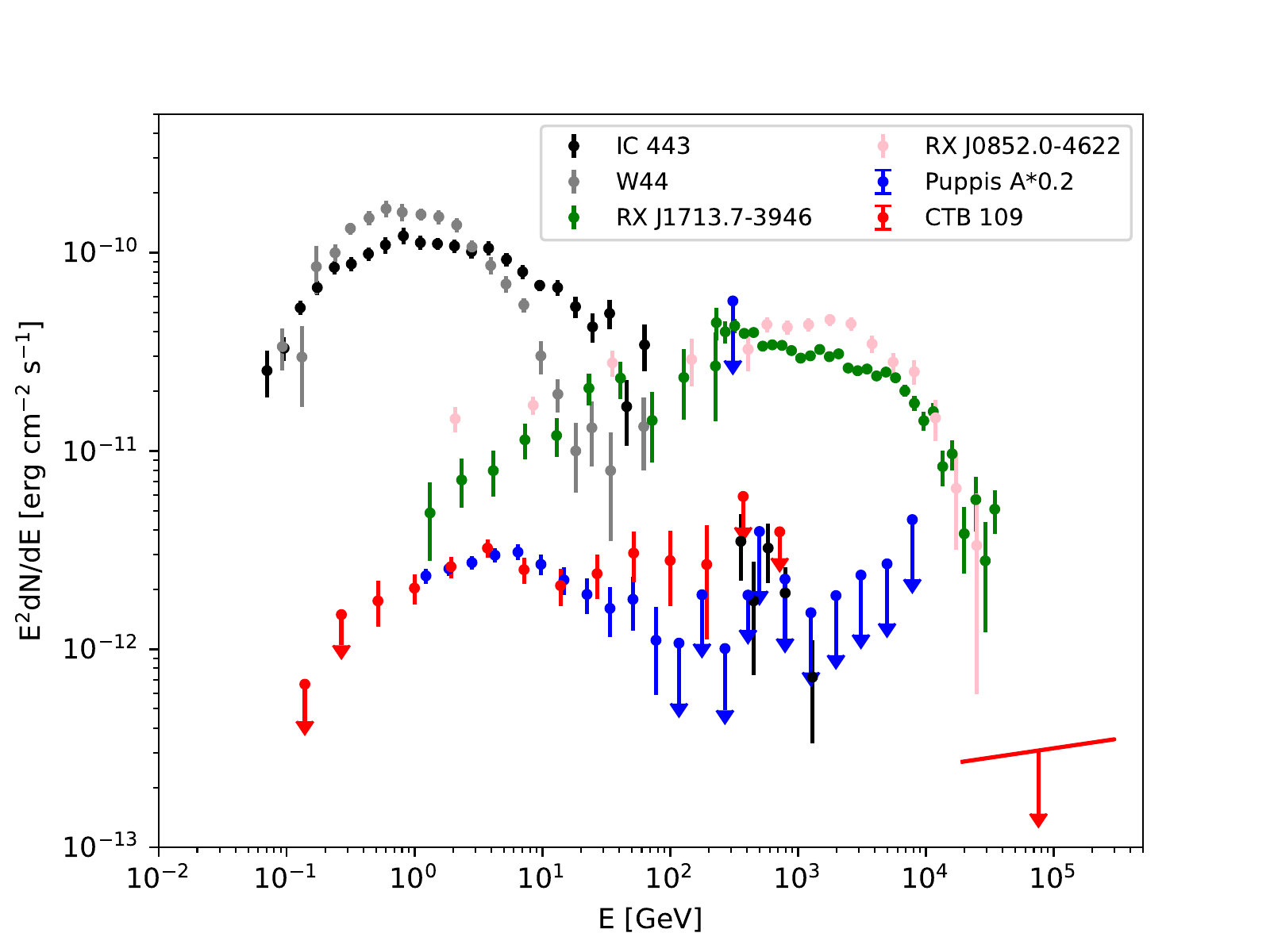}
        \includegraphics[width=0.49\textwidth]{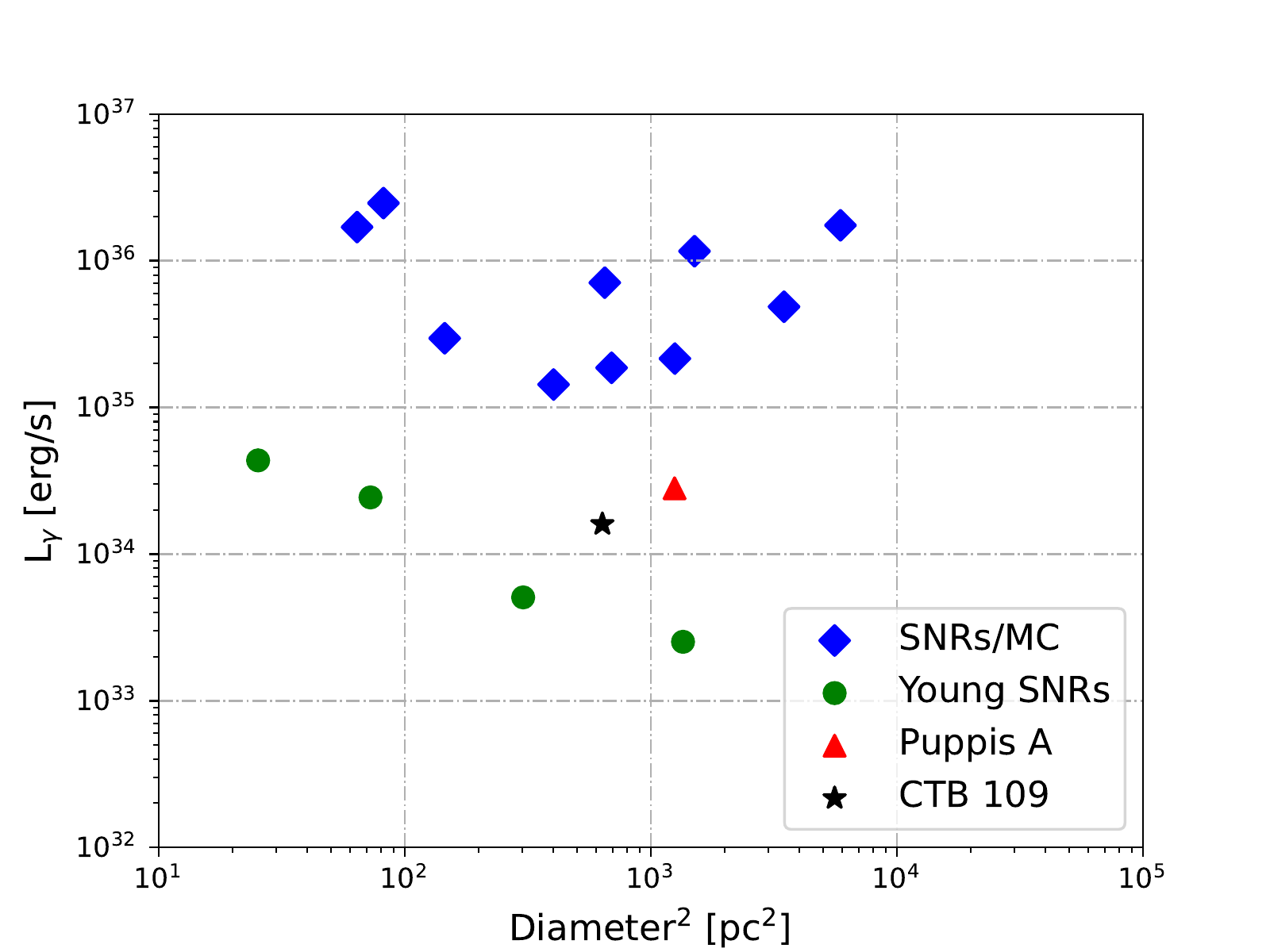}
	\caption{Left: Typical $\gamma$-ray SEDs of several prominent SNRs and CTB 109. The energy fluxes of Puppis A are scaled downward by five times for comparison. Right: 0.1 - 100 GeV $\gamma$-ray luminosities of several {\em Fermi}-LAT observed SNRs plotted against their diameter squared, which is reproduced from \citet{2012APh....39...22T} and \citet{2020AA...643A..28D} with the addition of CTB 109. The blue diamonds and green circles represent the old-aged SNRs interacting with molecular clouds and the young-aged SNRs. The red triangle shows SNR Puppis A, and CTB 109 is marked as the black star.}
	\label{fig3:multiSNRs}
\end{figure*}

In the left panel of Figure \ref{fig3:multiSNRs}, we plot the $\gamma$-ray spectra of several {\em Fermi}-LAT observed SNRs.
The old-aged SNRs, like IC 443 \citep[age $\sim$ 20 kyr;][]{2008AJ....135..796L, 2013Sci...339..807A} and W44 \citep[age $\sim$ 20 kyr;][]{1991ApJ...372L..99W, 2013Sci...339..807A}, show the spectral break at $\sim$ GeV, which makes their $\gamma$-ray emission brighter in the GeV band than in the TeV band \citep{2015ARNPS..65..245F}.
These SNRs are usually suggested to be interacting with MCs, and the $\gamma$-ray emission from such SNR/MC systems are suggested to be hadronic origin.
Another class of SNRs, including RX J1713.7-3946 \citep[age $\sim$ 1.6 kyr;][]{1997AA...318L..59W, 2018A&A...612A...6H} and RX J0852.0-4622 \citep[age $\sim$ 1.7-4.3 kyr;][]{2008ApJ...678L..35K, 2018A&A...612A...7H}, are young-aged systems with strong non-thermal X-ray emission and hard GeV spectra. The $\gamma$-ray emission from these SNRs are suggested to be from inverse Compton scattering of accelerated electrons (leptonic model), which shows the spectral curvature at $\sim$ TeV. 
The spectral analysis in Section \ref{results} shows an evident spectral curvature at $\sim$ several GeV for the $\gamma$-ray spectrum of CTB 109, which is similar to that of Puppis A \citep[age $\sim$ 4.45 kyr;][]{2012ApJ...755..141B, 2017ApJ...843...90X}.
In addition, the right panel of Figure \ref{fig3:multiSNRs} depicts the $\gamma$-ray luminosity from 100 MeV to 100 GeV as a function of the diameter squared for several {\em Fermi}-LAT detected SNRs. The data points of the sample are cited from \citet{2012APh....39...22T} and \citet{2020AA...643A..28D}, whose names, ages and classes are listed in Table \ref{table:SNRs}.
And the squared diameter could also be regarded as a reasonable indicator of SNR ages.
The $\gamma$-ray luminosity of CTB 109 is calculated to be 1.6 $\rm \times 10^{34}$ $\rm erg$ $\rm s^{-1}$ in the energy range of 100 MeV - 100 GeV.
The different $\gamma$-ray spectra of the different type of SNRs and the luminosity-diameter squared relation make CTB 109 and Puppis A to be distinguished both from several old-aged SNRs interacting with molecular clouds (e.g. IC443) and the young-aged SNRs with hard GeV $\gamma$-ray spectra (e.g. RX J1713.7-3946).


\begin{table}[!htb]
\centering
\caption {Ages of different type of SNRs}
\begin{tabular}{ccccccc}
\hline \hline
Class   & Name        &  Age (kyr)      &  Reference \\
\hline
SNRs/MC  & & & \\
 &  IC 443      &  $\sim$20       &   \citet{2008AJ....135..796L}   \\
 &  W44         &  $\sim$20       &   \citet{1991ApJ...372L..99W}   \\
 &  W51C        &  $\sim$30       &   \citet{1995ApJ...447..211K} \\
 &  W28         &  $\sim$33       &   \citet{2002AJ....124.2145V} \\
 &  G8.7-0.1    &  $\sim$15       &   \citet{1990Natur.343..146K} \\
 &  CTB 37A     &  $\sim$24       &   \citet{2014PASJ...66....2Y} \\
 &  W49B        &  5-6      &   \citet{2018AA...615A.150Z} \\
 &  3C391       &  $\sim$8.7      &   \citet{2018ApJ...866....9L} \\
 &  G349.7+0.2  &  $\sim$1.8      &   \citet{2014ApJ...783L...2T} \\
\hline
Young  SNRs  & & & \\
 &  RX J1713.7-3946      &  $\sim$1.6      &  \citet{1997AA...318L..59W}  \\
 &  RX J0852.0-4622      & 1.7-4.3   &  \citet{2008ApJ...678L..35K}  \\
 &  CasA                 &  0.35     &  \citet{2001AJ....122..297T} \\ 
 &  Tycho                &  0.45     &  \citet{1945ApJ...102..309B} \\
 &  G150.3+4.5           &  $\sim$1.0      &  \citet{2020AA...643A..28D} \\
\hline
 &  Puppis A    &  $\sim$4.45        &   \citet{2012ApJ...755..141B}  \\
 &  CTB 109     &  $\sim$9.0         &   \citet{2018MNRAS.473.1705S}  \\
\hline
\hline
\end{tabular} 
\label{table:SNRs}
\end{table}  

To explore the origin of the $\gamma$-ray emission from CTB 109, the leptonic and hadronic models are adopted to fit the multi-wavelength observations.
Different from the CR-hydro-NEI model adopted in \citet{2012ApJ...756...88C} including the SNR hydrodynamics, a semi-analysis calculation of nonlinear diffuse shock acceleration and the nonequilibrium ionization conditions behind the forward shock, here we adopted a phenomenological spectra for electrons and protons to constrain the distributions of cosmic rays, together considering the upper limit in the TeV band from HAWC.
In the leptonic model, the inverse Compton scattering (ICS) or bremsstrahlung processes of relativistic electrons are considered. 
And for ICS process, the radiation field includes the cosmic microwave background (CMB) and the infrared component from interstellar dust and gas with T = 30K \& u = 1 eV $\rm cm^{-3}$ \citep{2006ApJ...648L..29P, 2008ApJ...682..400P, 2019ApJ...874...50Z}.
The distance of CTB 109 is 3.1 kpc \citep{2018MNRAS.473.1705S}, and the density of the ambient medium is adopted to be n$_{\rm gas}$ = 1.1 cm$^{\rm -3}$ \citep{2012ApJ...756...88C}.
We use the {\em naima} package to fit the multi-wavelength data with the Markov Chain Monte Carlo (MCMC) algorithm \citep{2015ICRC...34..922Z}.

For the leptonic model, the spectra of electrons was first assumed to be a single power-law with an exponential cutoff (PL) in the form of 
\begin{equation}
    \frac{dN_{\rm e}}{dE} \propto \left(\dfrac{E}{E_0}\right)^{-\alpha_{e}} exp\left(-\dfrac{E}{E_{\rm cut,e}}\right) 
\end{equation}
And considering the GeV spectral curvature of the $\gamma$-ray emission, we also tested another electron distribution of a broken power-law model with an exponential cutoff \citep[BPL;][]{2019ApJ...874...50Z}, which follows:
\begin{equation}
    \frac{dN_{\rm e}}{dE} \propto exp\left(-\dfrac{E}{E_{\rm cut,e}}\right) 
    \begin{cases}
    \left(\dfrac{E}{E_0}\right)^{-\alpha_{e1}} \qquad \qquad \qquad \qquad ;  E < E_{\rm break}  \\
    \left(\dfrac{E_{\rm break}}{E_0}\right)^{\alpha_{e2}-\alpha_{e1}} \left(\dfrac{E}{E_0}\right)^{-\alpha_{e2}} \,; E \geq E_{\rm break}
    \end{cases}
\end{equation}
Here, $\alpha_{\rm e}$ and $\rm E_{cut,e}$ are the spectral index and the cutoff energy of electrons. 
$\rm E_{break,e}$ is the break energy, and the spectral indices below/above $\rm E_{break,e}$ are denotes as $\alpha_{\rm e1}$/$\alpha_{\rm e2}$.
To reduce the number of free parameters, the spectral index variation for the BPL model is set to be 1.0 ($\alpha_{\rm e2}$ = $\alpha_{\rm e1}$ + 1.0).
And the cutoff energy of electrons E$_{\rm cut,e}$ is calculated by equalling the synchrotron energy loss time-scale and the age of CTB 109, which is given by
E$_{\rm cut,e}$ = 1.25 $\times$ 10$^{\rm 7}$ t$_{\rm age; yr}^{-1}$ B$_{\rm \mu G}^{\rm -2}$ TeV. Here the age of CTB 109 is adopted to be  t$_{\rm age}$ = 9 kyr \citep{2018MNRAS.473.1705S}.

\begin{figure*}[!htb]
	\centering
    \includegraphics[width=0.48\textwidth]{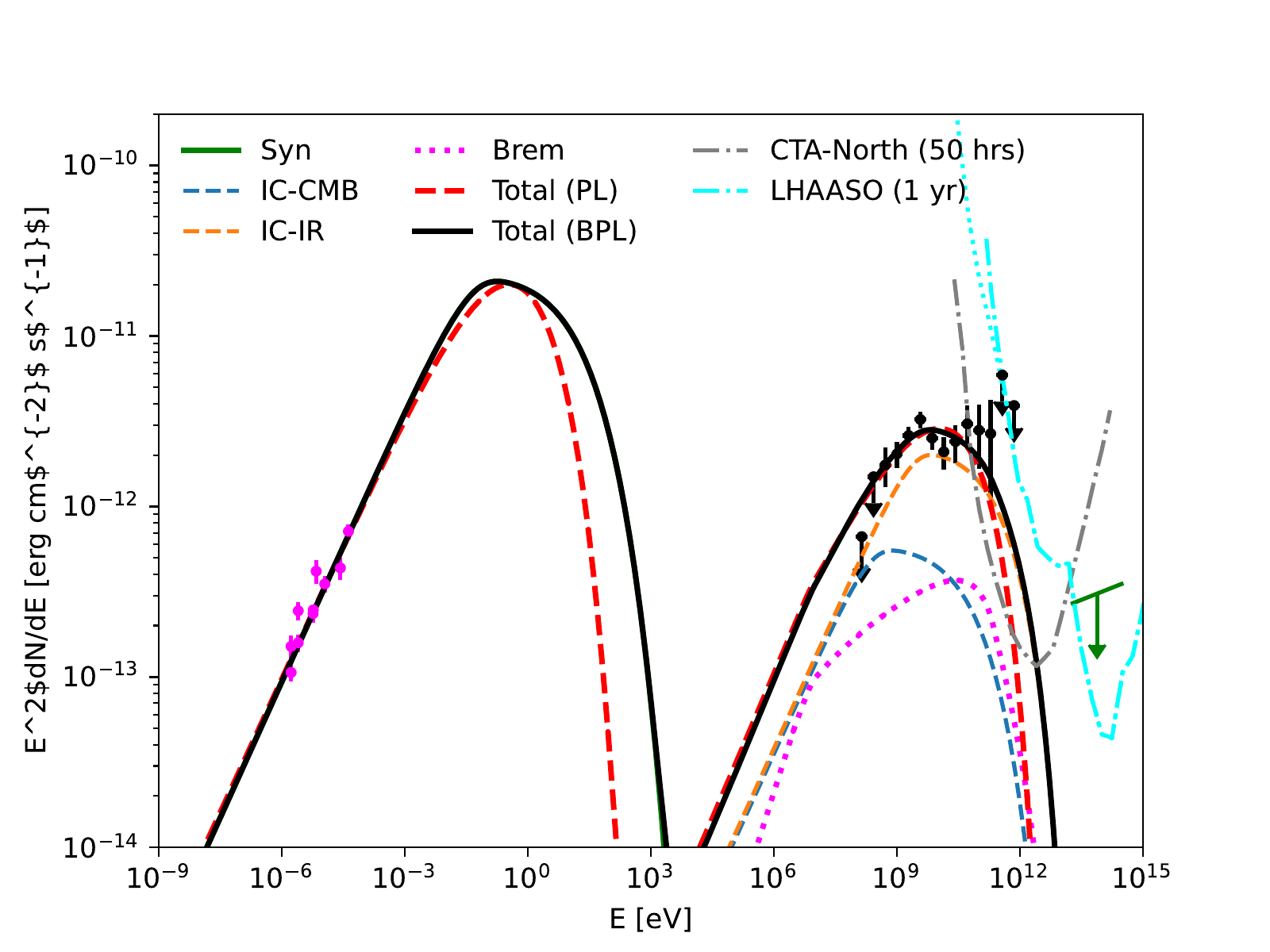}
    \includegraphics[width=0.48\textwidth]{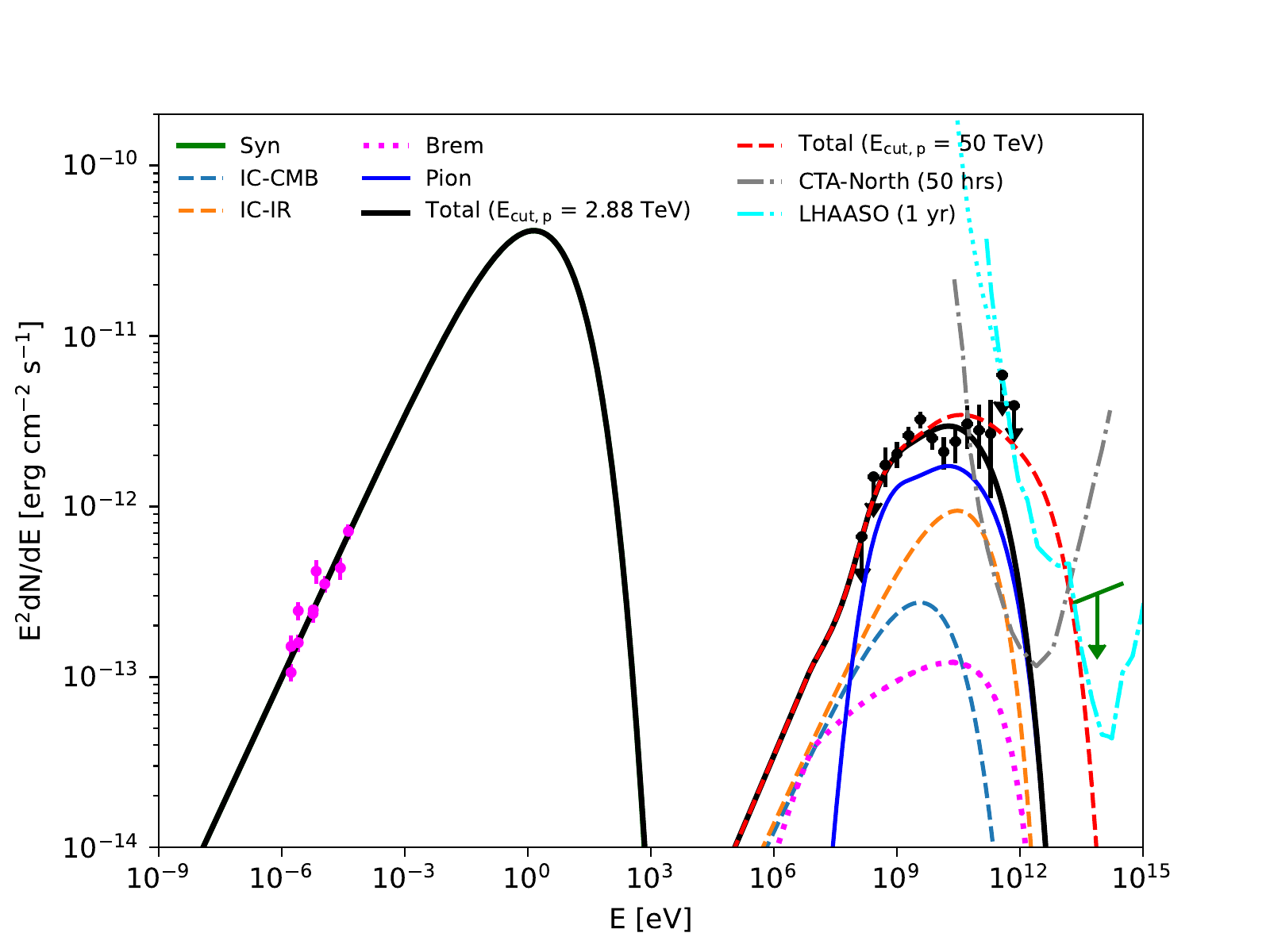}
\caption{Modeling of the multi-wavelength SED of CTB 109. 
And the radio data are from \citet{2006A&A...457.1081K}, together with the TeV upper limit by HAWC \citep{2019ICRC...36..674F}.
The left panel is for the leptonic model, where the PL and BPL models are presented by the black solid and red dashed lines, respectively. 
And the different radiation components for the BPL model are also overplotted, as the legend described.
The right panel is for the hadronic PL model, and the different radiation components are also illustrated in the legend.
The red dashed line is calculated with the cutoff energy of protons of E$_{\rm cut,p}$ = 50 TeV.
The cyan dotted and dotted–dashed lines represent the differential sensitivities of LHAASO (1 yr) with different sizes of photomultiplier tube \citep[PMT;][]{2019arXiv190502773C}. And the gray dot-dashed line shows the differential sensitivity of CTA-North \citep[50 hr;][]{2019scta.book.....C}.}
\label{fig:SEDmodel}
\end{figure*}

The best-fit parameters of leptonic models are shown in Table \ref{table:model}, and the corresponding modeled SED is given in the left panel of Figure \ref{fig:SEDmodel}.
For the leptonic PL model, the spectral index and cutoff energy of electrons are fitted to be 1.96 and 0.61 TeV, respectively.
A magnetic field strength of about 20 $\mu$G is needed to explain the flux in the radio band, 
which is similar to the value behind the shock fitted by \citet{2012ApJ...756...88C}. 
However, the total energy of electrons above 1 GeV of W$_{\rm e}$ = 1.31 $\times$ 10$^{\rm 48}$ erg is much lower than that required for the CR-hydro-NEI model in \citet{2012ApJ...756...88C}. 
The leptonic BPL model fitting gives the break energy of electrons to be $\sim$ 0.22 TeV, and other parameters are similar to that of the PL model.
Adopted the fitted magnetic field strength and the age of CTB 109, the cutoff energy of electrons is calculated to be about 3.44 TeV, which is larger than that of the PL model.

For the hadronic model, the spectral distributions of electrons and protons are both assumed to follow the PL model. And the spectral indices of electrons and protons are set to be equal to reduce the free parameters. Meanwhile, the ratio of the number of relativistic electrons to protons at 1 GeV, K$_{\rm ep}$, is assumed to be 0.01, which is in accord with the local measured cosmic ray abundance \citep{2012ApJ...761..133Y}.
The modeled SED is shown in the right panel of Figure \ref{fig:SEDmodel}, and the derived model parameters are listed in Table \ref{table:model}.
The spectral indices of particles are fitted to be 1.96.
The magnetic field strength of about 40 $\mu$G is two times larger than that in the leptonic model. And based on this value, the cutoff energy of electrons is calculated to be about 0.84 TeV.
However, the cutoff energy of protons, E$_{\rm cut,p}$, could not be well constrained due to the absence of the TeV $\gamma$-ray data for CTB 109.
With the spectral index of 1.96 for protons, we adjust the value of E$_{\rm cut,p}$ by manual to compute the $\gamma$-ray flux. 
And the upper limit in the TeV band by HAWC constrains the maximum of protons to be about 50 TeV. The corresponding modeled SED is shown as the red dashed line in the right panel of Figure \ref{fig:SEDmodel}.
The total energy of electrons and protons above 1 GeV are estimated to be W$_{\rm e}$ = 4.47 $\times$ 10$^{\rm 47}$ erg and  W$_{\rm p}$ = 5.52 $\times$ 10$^{\rm 49}$ (n$_{\rm gas}$/1.1 cm$^{\rm -3}$)$^{-1}$ erg, respectively. 
And such values are about one order lower than the fitted results in \citet{2012ApJ...756...88C}.

\begin{table}[!htb]
\centering
\caption {Parameters for Leptonic and Hadronic Models}
\begin{threeparttable}
\begin{tabular}{ccccccccc}
\hline \hline
Model   &  $\alpha_{\rm e/p}^{\,\,\,\,\,\,\,\text{a}}$
  &   E$_{\rm break,e}$   &  E$_{\rm cut,e}^{\,\,\,\,\,\,\,\,\,\,\text{b}}$  &  E$_{\rm cut,p}$   &  B     & W$_{\rm e}$  ($>$1 GeV)  &  W$_{\rm p}$ ($>$1 GeV)  \\
        &                                    &   TeV                 &  TeV              &  TeV               & $\mu$G & erg       & erg         \\
\hline
Leptonic & & & & & & & \\
PL  & 1.96$^{+0.07}_{-0.07}$  &  -- &  0.61$^{+0.20}_{-0.13}$ &  -- & 18.99$^{+3.26}_{-2.37}$ & 1.31$^{+0.15}_{-0.17} \times 10^{48}$  & -- \\
BPL  & 1.94$^{+0.07}_{-0.07}$  &  0.22$^{+0.07}_{-0.06}$ & 3.44 & -- & 20.10$^{+3.26}_{-3.12}$ & 1.29$^{+0.18}_{-0.15} \times 10^{48}$  & -- \\
\hline
Hadronic  & & & & & & & \\
PL  & 1.96$^{+0.05}_{-0.06}$  &  -- & 0.84 & 2.88$^{+15.78}_{-2.46}$ &  40.55$^{+4.44}_{-5.56}$ & 4.47$^{+0.97}_{-0.35} \times 10^{47}$  & 5.52$^{+1.17}_{-1.08} \times 10^{49}$ \\
\hline 
\end{tabular}
\begin{tablenotes}
\footnotesize
\item[a] For the leptonic BPL model, $\alpha_{\rm e}$ denotes the low-energy spectral index of electronic distribution, and the high-energy spectral index above E$_{\rm break,e}$ is $\alpha_{\rm e}$ + 1.0.
\item[b] For the leptonic BPL and hadronic models, E$_{\rm cut,e}$ is calculated by equalling the synchrotron energy loss time-scale and the age of SNR CTB 109.
\end{tablenotes}
\end{threeparttable}
\label{table:model}
\end{table}

The multi-wavelength SED fitting shown in Figure \ref{fig:SEDmodel} suggests that both the leptonic and hadronic model could explain the observational data of CTB 109. 
However, the low flux of the upper limit in the low $\gamma$-ray range and the center bright morphology of the $\gamma$-ray emission make the hadronic model favored.
Moreover, the different radiation models predict the different $\gamma$-ray fluxes in the TeV band, which will be uncovered with future high-resolution observations, such as Large High Altitude Air Shower Observatory \citep[LHAASO;][]{2019arXiv190502773C} and/or Cherenkov Telescope Array in the northern hemisphere\citep[CTA-North;][]{2019scta.book.....C}.

For the $\gamma$-ray emission of Sh 2-152, its spectrum is similar to that of other star forming regions (SFRs) or young massive star clusters (YMCs) detected by {\em Fermi}-LAT, such as W40 \citep{2020A&A...639A..80S}, W43 \citep{2020A&A...640A..60Y}, and NGC 6618 \citep{2022MNRAS.513.4747L}, etc.
More and more observations support SFRs/YMCs to be an important class of factories of Galactic cosmic rays \citep{2019NatAs...3..561A}.
In SFRs/YMCs, there are many young massive stars and core-collapse supernovas. And cosmic rays are expected to be efficiently accelerated by strong fast winds of young massive stars and shocks of SNRs. Meanwhile, SFRs host the dense molecular gas to drive star formation, and the $\gamma$-ray emission from them favor the hadronic interaction between the accelerated cosmic rays and dense molecular clouds around them \citep{2020SSRv..216...42B}.
In the hadronic scenario, we adopt a simple power-law spectrum for protons to fit the $\gamma$-ray SED of Sh 2-152, which is shown as the red solid line in the right panel of Figure \ref{fig2:sed}.
In \citet{2011A&A...535A...8R}, the diameter and total mass of the cluster of Sh 2-152 are determined to be $\gtrsim$ 4.0 pc and (2.45 $\pm$ 0.79)$\rm \times 10^{3}$ $M_{\odot}$, respectively.
And the average density of the ambient gas is estimated to be $\sim \rm 3 \times 10^{3}$ $\rm cm^{-3}$.
With the spectral index of 2.5 for protons, the total energy of protons above 1 GeV, W$_{\rm p}$, is calculated to be 2.17$\rm \times 10^{46} (n_{gas}/3000\;cm^{-3})\; erg$ to fit the $\gamma$-ray data of Sh 2-152.
The required total energy of protons of Sh 2-152 is orders of magnitude lower than that of other SFRs/YMCs \citep{2019NatAs...3..561A}, 
which could be attributed to the less identified OB stars and lower mass than the other detected systems \citep{2020A&A...639A..80S}.

\section{Conclusions}
\label{con}

In this work, we reanalyze the $\gamma$-ray emission from SNR CTB 109 using 13 years Pass 8 {\em Fermi}-LAT data. 
The center bright $\gamma$-ray morphology of CTB 109 is well consistent with its thermal X-ray emission, but is not consistent with the shell-type structure in the radio band.
The GeV $\gamma$-ray spectrum of CTB 109 described by a log-parabola model shows an evident spectral curvature at $\sim$ several GeV, which can naturally explain the upper limit derived from HAWC observations in the TeV band.
By comparing CTB 109 and other typical SNRs, the unusual $\gamma$-ray spectrum of CTB 109 and the luminosity-diameter squared relation 
make CTB 109 to be distinguished both from the young-aged SNRs with hard GeV $\gamma$-ray spectra and several old-aged SNRs interacting with molecular clouds.
The multi-wavelength data can be well explained with the leptonic or hadronic model.
However, considering the low flux of the upper limit in the low $\gamma$-ray range, together with the $\gamma$-ray morphology and the spectral curvature of the GeV spectrum, the hadronic model is favored, like SNR Puppis A.
The decisive evidence of the hadronic model may rely on the high-resolution observations in the TeV band by LHAASO and/or CTA in the future.

\acknowledgments

We would like to thank the anonymous referee for very helpful comments, which help to improve the paper. 
This work is supported by the Natural Science Foundation for Young Scholars of Sichuan Province, China (No. 2022NSFSC1808),
the Science and Technology Department of Sichuan Province (No. 2021YFSY0031, No.2020YFSY0016),
the Fundamental Research Funds for the Central Universities (No. 2682021CX073, No. 2682021CX074, No. 2682022ZTPY013), 
and the National Natural Science Foundation of China under the grants 12103040 and 12147208.

%



\bibliography{sample63}{}
\bibliographystyle{aasjournal}


\end{document}